# Untrue.News: A New Search Engine For Fake Stories

Vinicius Woloszyn[1], Felipe Schaeffer[2], Beliza Boniatti[2], Eduardo Cortes[3],
Salar Mohtaj[1,4], Sebastian Möller[1,4]

[1]Technische Universität Berlin, Germany
[2]Universitat de València, Spain
[3]Federal University of Rio Grande do Sul, Porto Alegre, Brazil
[4]German Research Centre for Artificial Intelligence (DFKI), Berlin, Germany

{woloszyn, salar.mohtaj, sebastian.moeller}@tu-berlin.de
{fesne, bebodos}@alumi.uv.es
egcortes@inf.ufrgs.br

## Abstract

In this paper, we demonstrate **Untrue News**, a new search engine for fake stories. Untrue News is easy to use and offers useful features such as: a) a multi-language option combining fake stories from different countries and languages around the same subject or person; b) an user privacy protector, avoiding the filter bubble by employing a bias-free ranking scheme; and c) a collaborative platform that fosters the development of new tools for fighting disinformation. Untrue News relies on Elasticsearch, a new scalable analytic search engine based on the Lucene library that provides near real-time results. We demonstrate two key scenarios: the first related to a politician â looking how the categories are shown for different types of fake stories â and a second related to a refugee â showing the multilingual tool. A prototype of Untrue News is accessible via **http://untrue.news**.

## 1. Introduction

A Web Search Engine (WSE) is an Information Retrieval System used to discover web pages relevant to specific queries (Brin and Page, 2012). Multipurpose web search engines, such as Google, Yahoo, Bing among others, have been employed for retrieving a wide range of information from the Internet, ranging from news articles to scientific papers. Normally, search results are composed of thousands of items that are commonly ranked based on some criteria, for example: the combination of popularity and relevancy. Some WSEs also include the user's activity history in their ranking schema, creating personalized results.

Some studies(Pariser, 2011; Machado et al., 2018) have shown that those characteristics tend to lead users to a special type bias called Filter Bubble. Since this problem has been identified, other systems, like the web browser DuckDuckGo[1], have become popular with the pledge to avoid this problem by not tracking, or "bubbling", users. However, this measure was not enough to stop the proliferation of disinformation on the web. We see the lack of dedicated solutions as a barrier to the mitigation of disinformation. For example, when a person wants to make an exploratory search about fake stories involving a particular subject or a politician, the Multipurpose WSEs provides an array of unrelated information turning the search into a thorny chore for the user.

In this context of abundant information and disinformation we introduce Untrue News as a new type of search engine; designed exclusively for retrieving fake stories and their authors, filling an important gap left by the multipurpose tools. An essential aspect of it is the ability to enrich the results with articles in different languages whilst avoiding the "filter bubble" of personalized results. For instance, analyzing the fake stories about *Greta Thunberg* in Austria or in Brazil and retrieving the results in the user's language and from different sources. Also, our system can provide the veracity of each rumor recovered directly in the retrieval web interface.

In summary, the contribution of this project is as follows:

1. The first web search engine for fake news;

2. Open-source web crawler for gathering of fake articles on the Internet;

3. Natural Language Processing pipeline for data enrichment of fake articles in different languages, composed of automatic and semi-automatic strategies, e.g. text translation for multiple languages, recognition, and linking of entities (i.e. person, organizations, locations, and more) to structured open semantic knowledge bases (DBpedia);

4. An information retrieval strategy for avoiding the filter bubble of personalized results.

The rest of this paper is organized as follows: section 2. introduces the features, section 3. describes the design of our architecture, and section 4. gives further details about our knowledge base. on section 5. our workflow for collaboration is exposed, and on section 6. two scenarios are presented to illustrate the usability of the platform. Finally, section 7. summarizes our perspectives and final considerations.

## 2. Features

In addition to concerns about a classification that is coherent, rich, and accurate, Untrue News presents features that emerge from the social needs of the users. The features highlighted below show functionalities that not only differ from conventional web search engines, but also enhance the user's experience. These are: a) multilingual system for translation of different stories; b) protection of user's privacy; and c) the building of an academic friendly platform.

---
[1]https://duckduckgo.com/

### 2.1. Multi languages

What are the fake stories being spread in Austria or Brazil about *Greta Thunberg*? How are these stories being interpreted in those two countries that speak different languages? Untrue News combines fake stories in different languages around the same subject or person providing unique results on the user's own language. That is, it will not only select the most relevant information around a chosen topic in the appropriate language (the user's language), but it will actually translate the information into the user's own language whenever necessary. This is a key feature since many of the rumors at the national level are not reported abroad and thus, not translated â or at least not translated well enough to be coherent and match the exact words typed by the users.

### 2.2. Privacy

Untrue News distinguishes itself from other search engines by not profiling users; it shows everyone the same search results. User's privacy is very important to us, so Untrue News makes sure that searches are concealed and protected, whilst avoiding the filter bubble of personalized search results. It is now known how public opinion has been persuaded on polemic topics, such as the 2016 US elections and the Brexit referendum, by the profiling of users according to their perceived views(Ward, 2018). That is, users with more conservative views are shown more articles and stories (even if fake ones) that support these views; likewise, a more liberal-leaning users are shown stories that match these views. This tendency to search for, interpret or believe in information that confirms one's prior beliefs is known as Confirmation Bias(Nickerson, 1998). This method of political persuasion based on user-profiling creates a toxic social environment of polarised opinions and disinformation, which does not in any way contribute to healthy discussions on important issues. To the contrary, it opens the door for impersonation, manipulation, deception, distraction and intolerance. This is where Untrue News comes in, as an unbiased and independent fact-checking tool for the promotion of democracy and the truth that does not create "filter bubbles".

In practical terms, Untrue News will only ever collect the data it absolutely needs to in order to improve its service; it will not create any type of user profile based on their search history because this data will only be stored for a short period of time and with great care to protect users' privacy. Moreover, Untrue News does not use any external tracking tool to optimize its service, which means that users can rest assure their searches are not being tracked, but in addition, users are able to switch off all tracking, being able to browse completely anonymously in a private window. User's searches are encrypted in order to protect them from potential eavesdroppers, and in doing so, we can guarantee that there will never be anyone between Untrue News and its users watching their searches. Finally, Untrue News will never sell your data or your searches to advertising companies, this goes directly against our ethos and the purpose of the platform.

### 2.3. Academic Friendly

We envision Untrue News as an academic umbrella for students and researchers interested in investigating different branches of knowledge; for example, User Experience, Big Data Analytics, Open Data, Machine Learning, and Language Technologies. Additionally, the source code of **Untrue.news** [2] is available on the internet as an Open Source project, for fostering collaboration in the development of new tools for fighting disinformation.

The Open Source character is one of the underpinnings of this project. As a collaborative platform, Untrue News aims to build a community around itself which will enable it to grow collectively and thus, have a greater potential for overcoming the challenges a project of this magnitude will inevitably encounter; i.e.: the language barrier. Besides providing free tools, information, and advice, Untrue News compile a rich multilingual knowledge base that can be used for fostering other initiatives in fighting disinformation (Mohtaj et al., 2019).

## 3. System's Architecture

The overall architecture of Untrue News is illustrated in Figure 1 showing the system working mainly in two phases. The first one is the data collection from a list of reliable fact-checking agencies. This has the role of gathering new rumors, classifying their authenticity, structuring the data, and indexing it in an Information Retrieval System. The second phase has the role of creating the interaction between the platform and the users. This phase uses a framework that provides the user interface and also communicates with the Information Retrieval System in order to deliver the rumors' information about a query submitted to the system by the user.

1. **Web Crawler**. Normally, search engines maintain real-time information by running an algorithm commonly called 'web crawler'. We have designed a web crawler exclusively for finding information about fake stories. For each distinct source included in our whitelist of reliable websites, a template was carefully developed to navigate through the website and extract only the metadata that is publicly available, such as: the claim, date, source, amongst others.

2. **Data Enrichment** This comprises several natural language-processing tasks to enrich the collected content in order to enable faceted browsing in the context of the text, and to enhance search capabilities. For instance, we use an automatic machine translation that enables the user to overcome language barriers and to be informed about rumors in different languages. Furthermore, Untrue News employs Google translation API[3], as it has presented the best translations for this sort of documents. Lastly, we have employed named entity recognition and linking the documents to open semantic databases, like *DBpedia Spotlight* (Mendes et al., 2011) has been used[4].

---
[2]https://github.com/vwoloszyn/fake_engine
[3]https://cloud.google.com/translate/

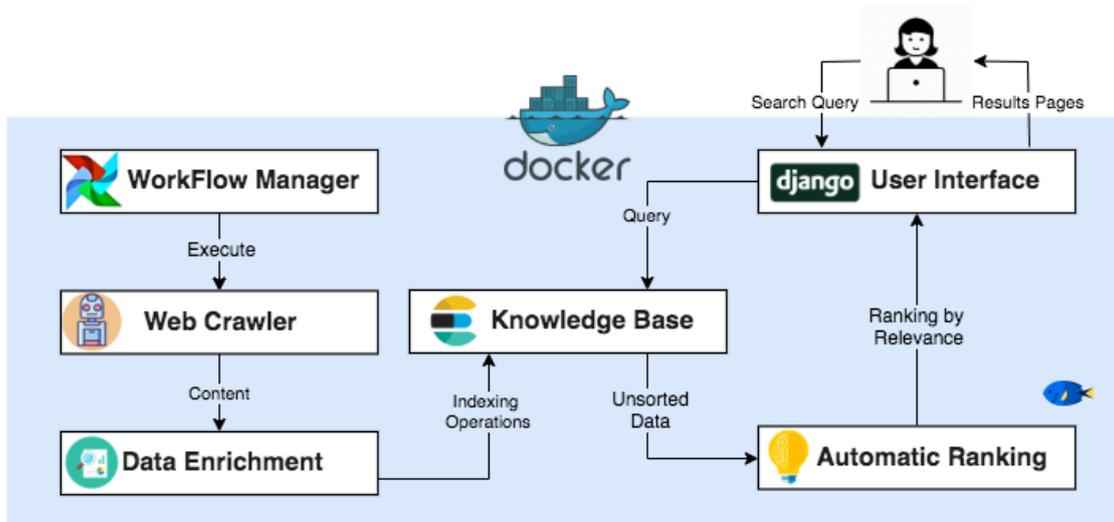

Figure 1: Untrue's Ecosystem

3. **Knowledge Base**. This is a technology used to store complex structured and unstructured information. We have used Elasticsearch to store our data, which is a scalable analytic search engine based on the Lucene library, that provides near real-time results. In our system, it is responsible for indexing all documents collected by the Web Crawler during the data collection phase, and retrieve them by using a schema-free JSON document, when requested during the User interface phase.

4. **Workflow Manager**. We have employed *Apache Airflow*[5] for orchestrating the data workflow in our system. It is an open-source platform that enables a system to programmatically schedule tasks and its dependencies as a Directed Acyclic Graphs (DAGs). In our system, it is employed for monitoring, alerting, and clustering the management of the data from the very early process of the data gathering to the indexing procedures into the Knowledge Base.

5. **User Interface**. The user interface, depicted in 2, has a role in managing the interaction with the user. We have employed the Django framework[6] since it follows the model-template-view architectural pattern that allows rapid and clear development. It is responsible for handling all the requests from users, making the connection to the database and generating HTML dynamically by using database-driven templates. The templates contain the static parts of the desired HTML output, as well as some special syntax describing how dynamically the content was obtained from the database and display all of it to the user.

6. **Automatic Ranking**. This plays a central role in many information retrieval systems, especially in search engines(Woloszyn and Nejdl, 2018). The main aim of a ranking is to optimally sort the result list, showing the most relevant results on the top and the least relevant on the bottom. Typically, learning to rank is a machine learning problem of predicting a score judgment, i.e.: how relevant or irrelevant a certain content is. Usually, to obtain this relevancy the judgment of humans are used as a benchmark for training machine learning models.

## 4. Knowledge Base

The Knowledge Base of rumors relies only in a highly reputable source or information. For instance, one valuable instrument to verify the truthfulness of a claim is 'fact-checking'. Usually, a fact-checking agency is a non-profit organization that provides independent investigations about questionable facts. We rely on sources considered by the fact-checking community as trustworthy according to International Fact-Checking Network's code of principles [7]. From the fact-checker, we use only metadata publicly available on the internet.

### 4.1. Rating Normalization

When publishers write an article, they can add an special web markup (schema.org/ClaimReview) to their articles containing structured data about the fact checked. For example, what was the claim being assessed, who made the claim, what was the verdict. The schema.org/ClaimReview markup establishes five different categories of the verdict: 1= "False", 2 = "Mostly false", 3 = "Half true", 4 = "Mostly true", 5 = "True". However, not all fact-checking agencies have adopted such standardization, being necessary a second step of normalization to make them comparable. In

---

[4] https://www.dbpedia-spotlight.org/
[5] https://airflow.apache.org/
[6] https://www.djangoproject.com/

[7] https://ifcncodeofprinciples.poynter.org/

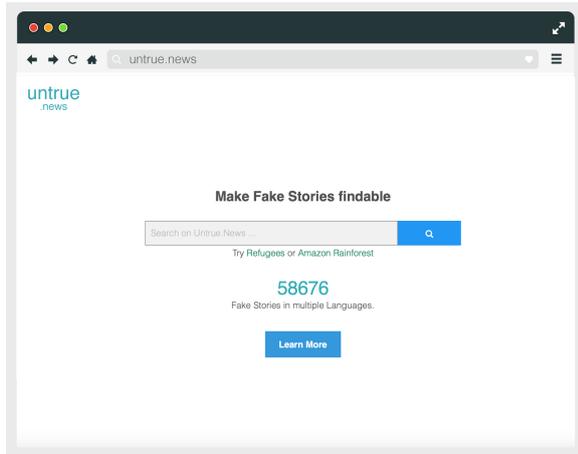

(a) Landing Page

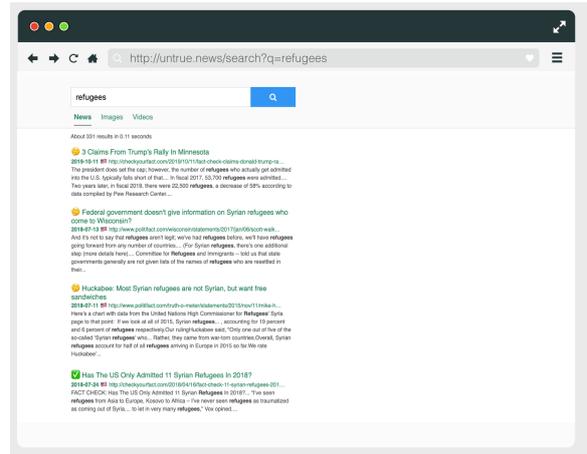

(b) Search Results

Figure 2: User Interface

order to tackle this issue, we have adopted 4 categories for data analysis (Tchechmedjiev et al., 2019):

- **TRUE**: statements completely accurate;
- **FALSE**: statements completely false;
- **MIXED**: statements partially accurate with some elements of falsity;
- **OTHER**: special articles that do not provide a clear verdict or do not match any other categories.

### 4.2. Statistics

Currently, the data collected from fact-checking agencies are composed of about 30.000 documents. This number is growing every day as new articles are published and the web crawler collects and indexes them. Table 1 presents the distribution of documents indexed by Untrue News according to the language. Most of the documents are obtained in English because it has a higher amount of news published everyday.

Figure 3 (a) presents the number of documents by year and language indexed by Untrue News and shows that our web crawler can extract English fact-checking reports published since 1995. However, for Portuguese and German, the documents obtained are from 2015 the earliest, since websites in those countries are newer. Finally, Figure 3 (b) presents the rating of distribution of these documents according to the classification of the 4 aforementioned categories.

We are constantly improving our Knowledge Base by adding new sources of information. For instance, our short-term goal is to include all fact-checking agencies listed as signatories from the International Fact-Checking Network[8].

## 5. Collaboration Workflow

There is an active collaboration between the academic community, students, and designers, working together to develop Untrue News. In this section we illustrate the workflow of collaboration between these groups, see Figure 4.

---

[8]https://ifcncodeofprinciples.poynter.org/signatories

Table 1: Fact-checkers indexed by Untrue News.

| URL | Language | # |
|---|---|---|
| fullfact.org<br>snopes.com<br>politifact.com<br>truthorfiction.com<br>checkyourfact.com | English | 26236 |
| piaui.folha.uol.com.br/lupa/<br>aosfatos.org/aos-fatos-e-noticia/<br>apublica.org/checagem/<br>g1.globo.com/e-ou-nao-e<br>e-farsas.com | Portuguese | 3473 |
| dpa.com<br>correctiv.org | German | 256 |

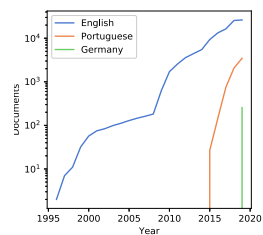

(a) Documents by year

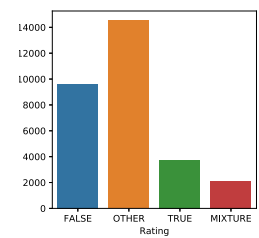

(b) Ratings

Figure 3: Knowledge Base Statistics.

As depicted in the figure 4, there are three different groups who play different roles in the system and push their coding to the GitHub of the project. There is also a project leader who merges requests and make changes on the master branch of the project. To achieve continuous integration in the project, the Jenkins triggers has been used to check any changes in the GitHub, pushing the changes in the server side as soon as any change is made in the repository.

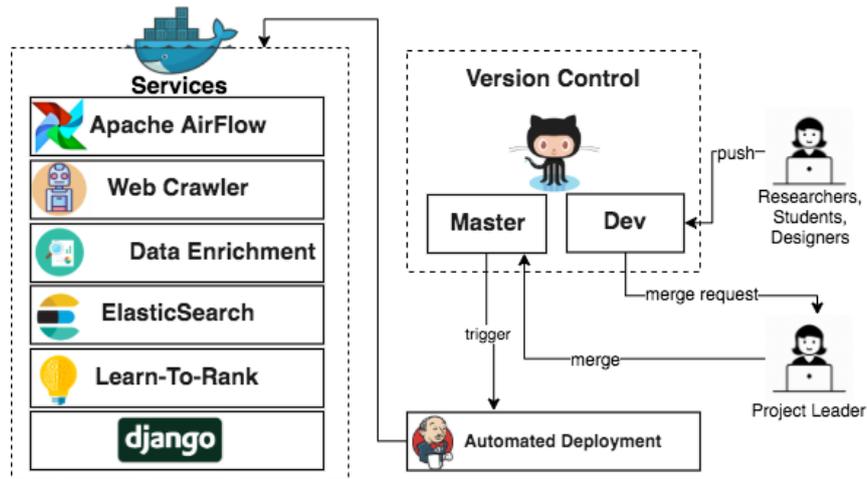

Figure 4: Collaboration Workflow

## 6. Demonstration Scenarios

The following two scenarios are cases that can be consulted and shown in the demo section of Untrue News. The first is regarding the search for a politician's name where the results can show rumors to be true, false or, most commonly, a mix of both. The second case highlights the multi-language functionality showing the search for a refugee-related issue that has been published and checked in English but can be read in the user's chosen language.

### 6.1. Searching for rumors related to a politician

Rumors about politicians have become a big strategy in election periods and the use of the internet for spreading false information, against or in favor of a candidate, can be decisive during that period. (Machado et al., 2018) reasons that fake news were decisive in the 2018 election of President Bolsonaro in Brazil. Compared to other candidates, "Bolsonaro supporters spread the widest range of known junk news sources" and their strategy of using fake news to attack opponents generated new statutory laws and investment in systems for fighting disinformation by the Brazilian government. The search for politicians on Untrue News results in different inquiries related to their names, which can be about rumors that: a) were spoken about them; b) they said about others; and c) that cites their names in the content. Accordingly, the results page shows the number of news that were found and how long the query of the database took. Each result of the list displays an icon related to the category information, the title of the article by the fact-checking agency, the date of the publication, the country in which the news was checked, a link to access the original fact, and an excerpt of it. The news that a Brazilian politician posted, about the transfer of 8 million dollars to the Ministry of Education, for example, will show with a symbol of indecision because it was confirmed by a fact-checking agency that it is true that these funds were transferred from the government, however, it involved other ministries besides education. Therefore, this story is classified as mixed because it involves true information â the transfer of federal funds â and false stories â which department the funds actually went to. In this case the user could go to the fact-checking agency's website to find out more about the details of the story.

### 6.2. Refugees checking out rumours in their mother tongue

The spread of fake stories on the Internet affects oppressed social groups the most (Peters, 2018). In the case of refugees, this happens mainly due to the process of cultural adaptation where barriers such as language and obliviousness of local laws and customs are grounds for conflict. Being aware of stories about themselves is an important element for improving their integration into society and the search for this kind of information on Untrue News can facilitate this process. Searching the keyword "refugees" on Untrue News will list a number of facts from various countries and written in an array of languages. The development of the multilingual tool enables the identification and translation of the listed content to be presented in the language chosen by the user; i.e.: Donald Trump's declaration saying that "Crime in Germany is up 10% plus since migrants were accepted". In the results list, Untrue News will show the title of the article with a red symbol beside it, symbolically representing the complete falseness of the information. It also shows the USA flag used to identify where the checked article was published, which can be used as an initial reference for the platform's multilingual system. The multilingual tool will allow the user to choose a language for the title and article to be translated to, regardless of the language in which they were published. Full translation functionality for fake stories is a work in process.

## 7. Final Considerations and Perspectives

In a time when rumors and false stories are decisive in building the collective thinking of society, Untrue News presents itself as a platform for fighting disinformation. As the first search engine for fake news, the platform has been developed to present orderly, relevant and accurate results, and its main features involve tools that enhance the user experience. Referring to the key features of the platform, the multilingual system for the identification and translation of results aims to aggregate users from different countries so that communication is offered to everyone equally. The users privacy is protected, providing consistent results but not standardized, whilst acting in the deconstruction of the filter bubbles. Finally, as an academic umbrella, the platform aims to enable other researchers and professionals to combat misinformation by working together based on true and checked information.

Untrue News uses open-source platforms such as a workflow manager to automate the tasks, an exclusive web crawler for collecting fake news, and a data enrichment process; all of them to build a consolidated knowledge base. The results showed through the user interface are ranked automatically by relevance and are organized by a collaborative workflow encompassing academy, students and designers. Currently, Untrue News is ranking about 30.000 documents between English, Portuguese and German. Since there is no standardization employed by the fact-checking agencies, Untrue News classifies the articles in four categories: true, false, mixed and other. This data normalization makes it possible to show on the search results page which category the story is in, providing a clear information for the user.

The perspectives for Untrue News are based on three points: a) technical amelioration in the searches; b) improvement in platform usability; and c) development of collaborative processes. The first concerns new types of searches such as searching for images and videos or searching for news-related entities. This improvement will bring new types of results with more comprehensive information about the stories, the agencies and the false images. The second deals with the user experience on Untrue News, both regarding visual aspects and accessibility of information. Usability tests are being developed to understand best practices for displaying information to each user, whether by lists, symbols or plots generated for each result. Finally, collaborative process development includes a content production section on important topics such as misinformation and false stories, as well as curatorial processes where collaborators will be able to suggest edits â especially for results in other languages. Building a collaborative, reliable and open-source platform for collective use is the main focus of Untrue News.

## 8. Bibliographical References